\documentstyle[12pt,epsf,here]{article}

\def \be {\begin{equation}}
\def \e {\end{equation}}
\def \bea {\begin{eqnarray}}
\def \ea {\end{eqnarray}}

\def \no {\nonumber}

\def \Vec#1{\mbox {\bf #1}}

\def \sub {\scriptscriptstyle}

\newcommand{\To}[2]{\stackrel{#1}{\hbox to #2 pt{\rightarrowfill}}}

\def \vector#1{\stackrel{\hspace{-0.45em}\longrightarrow}{#1}}

\def\np#1#2#3{{\it Nucl.~Phys.\/}~{\bf B#1} (19#2) #3}
\def\pl#1#2#3{{\it Phys.~Lett.\/}~{\bf B#1} (19#2) #3}

\def\prd#1#2#3{{\it Phys.~Rev.\/}~{\bf D#1} (19#2) #3}

\def\jpg#1#2#3{{\it J.~Phys.\/}~{\bf G#1} (19#2) #3}
\def\cpc#1#2#3{{\it Comput.~Phys.~Commun.\/}~{\bf #1} (19#2) #3}

\begin{document}  
\vspace*{-2cm}  
\renewcommand{\thefootnote}{\fnsymbol{footnote}}  
\begin{flushright}  
hep-ph/9903315\\
DTP/99/30\\  
March 1999\\  
\end{flushright}  
\vskip 65pt  
\begin{center}  
{\Large \bf Anomalous Quartic Couplings\\[3truemm] in \boldmath
$W^+W^-\gamma$,
\boldmath $Z^0Z^0\gamma$ and \boldmath $Z^0\gamma\gamma$ Production
\\[3truemm] 
at Present and Future \boldmath $e^+e^-$ Colliders} \\ 
\vspace{1.2cm} 
{\bf  

W.~James~Stirling${}^{1,2}$\footnote{W.J.Stirling@durham.ac.uk}  and
Anja Werthenbach${}^1$\footnote{Anja.Werthenbach@durham.ac.uk} }\\  
\vspace{10pt}  
{\sf 1) Department of Physics, University of Durham,  
Durham DH1 3LE, U.K.\\  
  
2) Department of Mathematical Sciences, University of Durham,  
Durham DH1 3LE, U.K.}  
  
\vspace{70pt}  
\begin{abstract}
The production of three electroweak gauge bosons in high-energy 
$e^+e^-$ collisions offers a window on  anomalous
quartic gauge boson couplings. We investigate the effect of
three possible anomalous couplings on the cross sections
for $W^+W^-\gamma$, $Z^0Z^0\gamma$ and $Z^0\gamma\gamma$ productions
at LEP2 ($\sqrt{s} = 200$~GeV) and at a future linear collider
($\sqrt{s} = 500$~GeV). We find that the combination of energies and
processes
provides  reasonable discrimination between the various anomalous
contributions.
\end{abstract}
\end{center}  
\vskip12pt

\setcounter{footnote}{0}  
\renewcommand{\thefootnote}{\arabic{footnote}}  
  
\vfill  
\clearpage  
\setcounter{page}{1}  
\pagestyle{plain} 
 
\section{Introduction} 

In the Standard Model (SM), the couplings of the gauge bosons and fermions
are tightly constrained by the requirements of gauge symmetry.
In the electroweak sector, for example, this leads to trilinear $VVV$ 
and quartic $VVVV$ interactions
between the gauge bosons $V=\gamma, Z^0, W^\pm$ with completely specified
couplings. Electroweak symmetry breaking via the Higgs mechanism gives
rise
to additional Higgs -- gauge boson interactions, again with specified
couplings.\\

The trilinear and quartic gauge boson couplings probe different 
aspects of the weak interactions. The trilinear couplings directly  test
the
non-Abelian gauge structure, and possible deviations from the SM
forms have been extensively studied
in the literature, see for example \cite{tgvtheory} and references therein. 
Experimental bounds have also been obtained \cite{tgvexpt}. 
In contrast, the quartic couplings 
can be regarded as a more direct window on electroweak symmetry breaking,
in particular to the scalar sector of the theory (see for example
\cite{godfrey}) or, 
more generally, on new physics which couples to electroweak bosons. \\

In this respect it is quite possible that the quartic couplings deviate
from their SM values 
while the triple gauge vertices do not. For example,
if the mechanism for electroweak symmetry breaking does not reveal itself
through 
the discovery of new particles such as the Higgs 
boson, supersymmetric particles or technipions it is  
possible that anomalous quartic couplings could provide the first evidence 
of new physics in  this sector of the electroweak theory \cite{godfrey}.
\\

High-energy colliders provide the natural environment for studying
anomalous quartic couplings.  The paradigm process is 
$f \bar f \to VVV$, with $f=e$ ($e^+e^-$ colliders) or $f=q$
(hadron-hadron colliders),
where one of the Feynman diagrams corresponds to $f \bar f \to V^* \to
VVV$. In this context,
one may consider the quartic-coupling diagram(s) as the signal, while the
remaining 
diagrams constituting the background. The sensitivity of a given process
to anomalous
quartic couplings depends on the relative importance of these two types of
contribution,
as we shall see. \\

In this study we shall focus on $e^+e^-$ collisions, and quantify the
dependence
of various $e^+e^- \to VVV$ cross sections on the anomalous couplings. We
shall
consider in particular $\sqrt{s} = 200$ and $500$~GeV, corresponding to
LEP2 and a future
linear collider (LC) respectively. For obvious kinematic reasons,
processes where at least
one of the gauge bosons is a photon have the largest cross sections.
Indeed, $VVV$ production
with $V=Z^0,W^\pm$ are kinematically forbidden at 200~GeV and
suppressed at 500~GeV.
We therefore consider $W^+W^-\gamma$, $Z^0 Z^0 \gamma$ and $Z^0
\gamma\gamma$ production.
Each of these contains at least one type of quartic
interaction.\footnote{We ignore the 
process $e^+e^- \to \gamma\gamma\gamma$ which involves no trilinear or
quartic interactions.} \\

There have been several studies of this type reported in the literature
\cite{belanger,ghadir}.
Our aim is partly to complete as well as update these, and partly to
assess the relative merits of the 
above-mentioned
processes in providing information on the anomalous couplings. 
Note that
our primary interest is in the so-called `genuine' anomalous quartic
couplings, i.e. those which give no contribution to the trilinear
vertices. \\

In the following section we review the various types of anomalous quartic
coupling
that might be expected in extensions of the SM. In Section~3 we present
numerical 
studies illustrating the impact of the anomalous couplings on various
$VVV$ cross sections.
Finally in Section~4 we present our conclusions.

\section{Anomalous gauge boson couplings}

The lowest dimension operators which lead to genuine quartic couplings 
where at least one photon is involved are of dimension 6 \cite{belanger}.
A
dimension 4 operator is not realised since a custodial $SU(2)$ symmetry is 
required to keep the $\rho$ parameter, $\rho = M_W^2/(M_Z^2 \cos ^2
\theta_w)$, 
close to its measured SM value of 1.
Thus the 4-dimensional operator
\bea
{\cal L}_4 = - \frac{1}{4} \, g^2 ( \vector{W _{\mu}} \times \vector{W
_{\nu}}) \,  
( \vector{W ^{\mu}} \times \vector{W ^{\nu}})
\ea 
with 
\bea
\label{wb}
\vector{W _{\mu}} = \left( \begin{array}{c}  \frac{1}{\sqrt{2}} (
W_{\mu}^+ 
+ W_{\mu}^-) \\  \frac{i}{\sqrt{2}} ( W_{\mu}^+ - W_{\mu}^-) 
\\  W_{\mu}^3 - \frac{g^{\prime}}{g} B_{\mu} \end{array} \right)
\ea
and
\bea
 W_{\mu}^3 - \frac{g^{\prime}}{g} B_{\mu} &=& \cos \theta_w\, 
 Z_{\mu} + \sin \theta_w A_{\mu} - \frac{e}{\cos \theta_w} 
 \frac{\sin \theta_w}{e} ( -\sin \theta_w Z_{\mu} +\cos \theta_w A_{\mu})
\no \\ 
&=& \frac{Z_{\mu}}{\cos \theta_w} \ . 
\ea
does not involve the  photon field $A_{\mu}$.
The other possible 4-dimensional operator \cite{belanger}
\bea
\widetilde{{\cal L}_4} = -ie \frac{\lambda _{\gamma}}{M_W^2}\, 
F^{\mu \nu} W_{\mu \alpha}^{\dagger} W^{\alpha}_{\phantom{\alpha} \nu}
\ea
with 
\bea
F^{\mu \nu}&=&  \partial _{\mu} A_{\nu} - \partial _{\nu} A_{\mu} \no \\
W_{\mu \nu}&=& \partial _{\mu} \Vec{W}_{\nu} - \partial _{\nu}
\Vec{W}_{\mu} 
-g \Vec{W}_{\mu} \times \Vec{W}_{\nu}
\ea
and
\bea
\label{wbsm}
\Vec{W}_{\mu} = \left( \begin{array}{c}  \frac{1}{\sqrt{2}} ( W_{\mu}^+ 
+ W_{\mu}^-) \\  \frac{i}{\sqrt{2}} ( W_{\mu}^+ - W_{\mu}^-) 
  \\  \cos \theta_w Z_{\mu} +\sin \theta_w A_{\mu} \end{array} \right) \ ,
\ea
generates trilinear couplings in addition to 
 quartic ones and is therefore not `genuine'. 
In Section 4 we will briefly discuss the impact of possible non-zero anomalous trilinear couplings on our analysis.\\
 
We are therefore left with several 6-dimensional operators. First  the
neutral
and the charged Lagrangians, both giving anomalous contributions
to the $VV\gamma\gamma$ vertex, with $VV$ either being $W^+W^-$ or
$Z^0Z^0$.
\bea
\label{L0}
{\cal L}_0 &=& - \frac{e^2}{16 \Lambda^2}\, a_0\, F^{\mu \nu} \, F_{\mu
\nu} \vector{W^{\alpha}} \cdot \vector{W_{\alpha}} \no \\
&=&  - \frac{e^2}{16 \Lambda^2}\, a_0\, \big[ - 2 (p_1 \cdot p_2 ) 
( A \cdot A) + 2 (p_1 \cdot A)(p_2 \cdot A)\big] \no \\
&& \hspace{1.5cm} {\sub \times} \big[ 2 ( W^+ \cdot W^-) +  (Z \cdot Z) /
\cos ^2 \theta_w \big]  \quad ,
\ea
\bea
\label{Lc}
{\cal L}_c &=& - \frac{e^2}{16 \Lambda^2}\, a_c\, F^{\mu \alpha} \, F_{\mu
\beta} \vector{W^{\beta}} \cdot \vector{W_{\alpha}} \no \\
&=& - \frac{e^2}{16 \Lambda^2}\, a_c\, \big[- (p_1 \cdot p_2)\, A^{\alpha}
A_{\beta} +(p_1 \cdot A)\, A^{\alpha} p_{2 \beta} \big. \no \\
&& \hspace{1.8cm}\big. \quad \quad + (p_2 \cdot A)\,
p_1^{\alpha} A_{\beta} -(A \cdot A)\, p_1^{\alpha} p_{2 \beta} \big] \no
\\
&& \hspace*{1.5cm} {\sub \times} \big[ W_{\alpha}^- W^{+ \beta} +
W_{\alpha}^+ W^{-
\beta} + Z_{\alpha} Z^{\beta} / {\cos ^2 \theta_w} \big] \ . 
\ea
where $p_1$ and $p_2$ are the photon momenta.\\

Since we are interested in the anomalous $VV\gamma\gamma$ contribution 
we pick up the corresponding part of the Lagrangian. To 
obtain the Feynman rules for the corresponding
 vertex (in agreement with \cite{eboli}) 
we have to multiply by 2 for the two identical photons (as well as for the
$Z^0$s in the case of $VV=Z^0Z^0$)
and by $i$ for convention. \\

Finally, an anomalous $WWZ\gamma$ vertex is obtained from the Lagrangian
\bea
\label{Ln}
{\cal L}_n &=&  - \frac{e^2}{16 \Lambda^2}\, a_n \epsilon_{ijk} W_{\mu
\alpha}^{(i)} W_{\nu}^{(j)} W^{(k)\alpha} F^{\mu \nu} \no \\
&=&  - \frac{e^2}{16 \Lambda^2 \cos \theta_w} \, a_n \, \big( p^{ \nu}
A^{\mu} - 
p^{\mu} A^{\nu} \big) \no \\
&& {\sub \times}  \left( - W^-_{\nu} p_{\mu}^+\,  ( Z \cdot W^+) +
W_{\nu}^+
p_{\mu}^-\, ( Z \cdot W^-) + Z_{\nu} p_{\mu}^+\, (W^+ \cdot W^-) \right.
\no \\
&& - Z_{\nu} p_{\mu}^-\, (W^+ \cdot W^-) + W_{\nu}^- W_{\mu}^+\, (p^+
\cdot Z) - W_{\nu}^+ W_{\mu}^- \,(p^-  \cdot Z) \no \\
&& - Z_{\nu} W_{\mu}^+\, (p^+ \cdot W^-) + Z_{\nu}W_{\mu}^-\, ( p^- \cdot
W^+) - W^+_{\nu} p^0_{\mu} \,(Z \cdot W^-) \no \\
&& \left. + W_{\nu}^- p^0_{\mu} \, (Z \cdot W^+) - W_{\nu}^- Z_{\mu}\,
(p^0 \cdot W^+) +W_{\nu}^+ Z_{\mu}\, (p^0 \cdot W^-) \right) 
\ea
where $W_{\nu}^{(j)}$ are the components of the vector (\ref{wb}) and $p,
p^+, p^-$ and $p^0$ are the momenta of the photon, the $W^+$, the $W^-$
and the $Z^0$ respectively.\\

It follows from the Feynman rules  that any
anomalous contribution is {\it linear} in the photon energy 
$E_{\gamma}$. This means that it is the hard tail of the
photon energy distribution that is most affected by the anomalous
contributions, but unfortunately the cross sections here are very small.
In the following numerical studies we will impose a lower energy photon
cut of $E_{\gamma}^{\rm min} = 20$~GeV. Similarly, there is also
no anomalous contribution to the initial state photon radiation, and
so the effects are largest for centrally-produced photons. We therefore
impose an additional cut of $\vert \eta_\gamma\vert <
2$.\footnote{Obviously 
in practice these cuts will also be tuned to the detector capabilities.}
\\

A further consideration concerns the effects of beam polarisation.
One of the `background'   (i.e. non-anomalous) diagrams for $e^+e^-
\to W^+W^-\gamma$ production is where all three gauge bosons are attached
to the electron line. Such contributions can be suppressed by an
appropriate
choice of beam polarisation (i.e. right-handed electrons)
 thus enhancing the anomalous signal. We will illustrate this below. \\

Finally, the anomalous parameter $\Lambda$ that appears in all the
above anomalous contributions has to be fixed. 
In practice, $\Lambda$ can only be meaningfully specified in the context of a specific model for the new physics giving rise to the quartic couplings. One example is an excited $W$ scenario $W^+ \gamma \to  W^* \to  W^+ \gamma$, where we would expect $\Lambda \sim  M_{W^*}$
and $a_i$ to be related to the decay width for $W^* \to  W + \gamma$.
However, in order to make our analysis independent of any such model, we choose to fix $\Lambda$ at a reference
value of $M_W$, following the conventions adopted in the literature. Any other choice of $\Lambda$ (e.g. $\Lambda = 1$~TeV) 
results in a trivial rescaling of the anomalous parameters $a_0$, $a_c$
and $a_n$.\\

\section{Numerical studies}

In this section we study the dependence of the cross sections on the 
three anomalous couplings defined in Section~2. 
As already stated, we apply a cut on the photon 
energy $E_{\gamma} > 20$~GeV to take care of the infrared singularity,
 and a cut on the photon rapidity 
 $|\eta_{\gamma}| < 2$ to avoid collinear singularities. We do not include
any branching ratios
 or acceptance cuts on the decay products of the produced $W^\pm$ and
$Z^0$ bosons, since we assume
 that at $e^+e^-$ colliders the efficiency for detecting these is high. \\
 
\begin{figure}[H]
\centerline{\epsfysize=17cm\epsffile{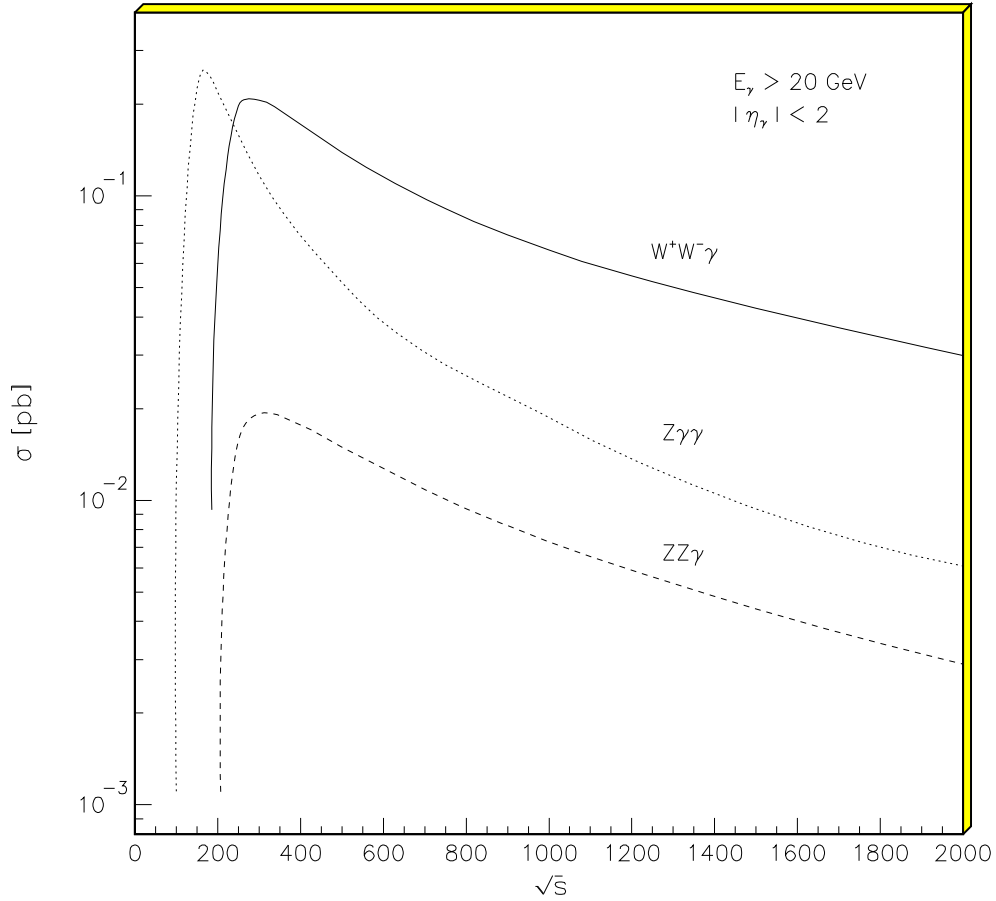}}
\vspace{-2cm}
\caption{\label{total}{Total SM cross sections for
$e^+e^- \to  W^+W^-\gamma,\; Z^0Z^0\gamma,\;  Z^0\gamma\gamma$ (in pb)
 as a function of $\protect\sqrt{s}$.}}
\end{figure}

We first consider the SM cross sections for the processes of interest,
i.e.
with all anomalous couplings set to zero. Figure~\ref{total} 
shows the collider energy dependence
of the $e^+e^- \to W^+W^-\gamma$,  $ e^+e^- \to Z^0Z^0\gamma$ 
and $e^+e^- \to Z^0\gamma\gamma$ cross sections.\footnote{Note that
although these
cross sections have appeared before in the literature, we are unable 
to reproduce the results for $\sigma(Z\gamma\gamma)$ given in Figure~2 of
Ref.~\cite{barger}. To cross check our results we used MADGRAPH
\cite{madgraph}.}
\vspace{-1.5cm}

\begin{figure}[H]
\centerline{\epsfysize=17cm\epsffile{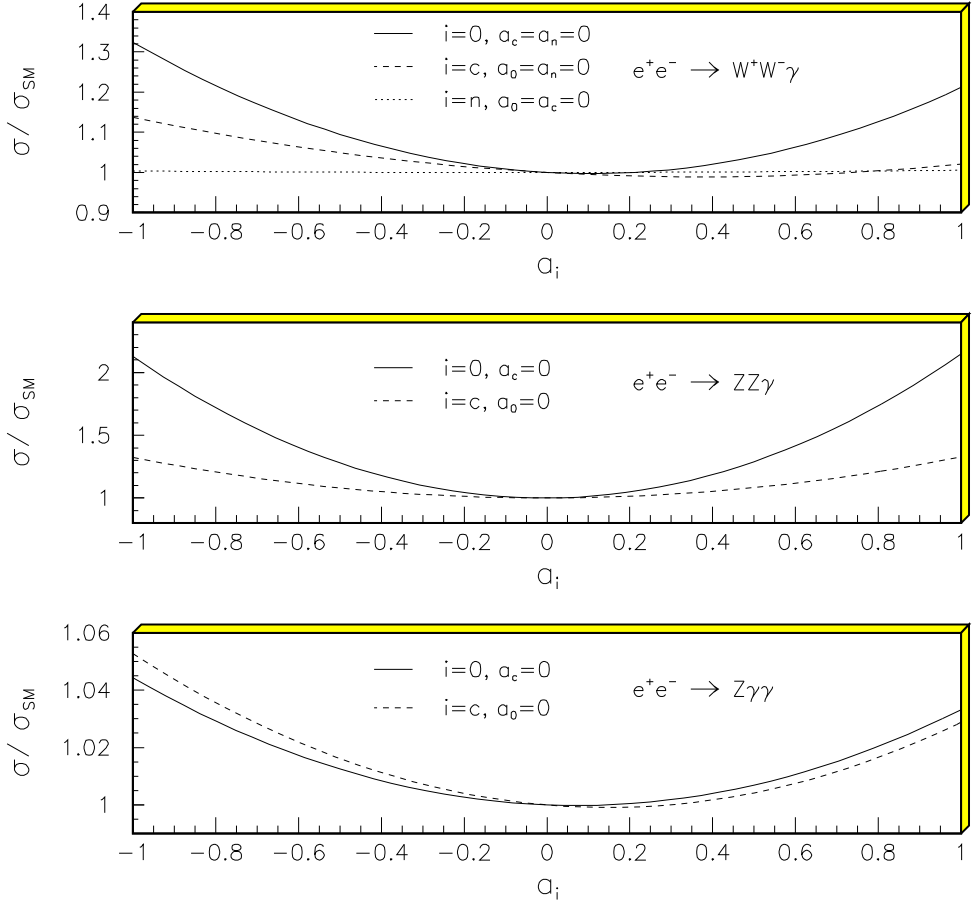}}
\vspace{-2cm}
\caption{\label{asubi}{Influence of the anomalous parameters 
on the total cross sections, normalised to their SM 
values, at $\sqrt{s}= 500 $~GeV.}}
\end{figure}

Next we study the influence of each of the three anomalous 
parameters $a_0, a_c$ and $a_n $ separately in order to gauge
the impact of each on the cross section. Note that $\sigma(W^+W^-\gamma)$
depends on all 
three parameters, while $\sigma(Z^0Z^0\gamma)$ and
$\sigma(Z^0\gamma\gamma)$
depend only on $a_0$ and $a_c$. 
Figure~\ref{asubi} shows the dependence of the three total 
cross sections of Figure~\ref{total} at $\sqrt{s} = 500$~GeV 
on the anomalous parameters.
In each case the cross section is normalised to its SM value,
and the cuts are the same as in Figure~\ref{total}.
\vspace{-1.5cm}
As expected the dependence on the $a_i$ is  quadratic,  since they 
appear linearly in the matrix element. The fact that the 
minimum of the curves is close to the SM point $a_i=0$ shows that the
interference between the anomalous and standard parts of the matrix 
element is small.
The anomalous parameters have a markedly different effect
on the three cross sections. 
Evidently  $a_0$ has the 
largest influence, particularly on $\sigma(Z^0Z^0\gamma)$.
The reason for this is easily understood. The anomalous
process $e^+e^- \to \gamma^* \to Z^0Z^0\gamma$ has a much larger impact
on $\sigma(Z^0Z^0\gamma)$ since there are only six other SM diagrams.
In contrast, $e^+e^- \to \gamma^* \to W^+W^-\gamma$ has a much larger
SM `background' set of diagrams to contend with. Note also that the
anomalous 
contributions are enhanced by a 
 factor $1/ \cos ^4  \theta_w$ compared to the $WW\gamma \gamma$ 
 vertex.\\

Of course the important question is which of the three processes
offers the best chance of detecting an anomalous quartic coupling
at a given collider energy.
To answer this we need to combine the information from 
Figs.~\ref{total} and \ref{asubi} to see whether enhanced sensitivity
can overcome a smaller overall event rate. We also need to consider 
{\it correlations} between different anomalous contributions 
to the same cross section.\\

\begin{figure}[H]
\centerline{\epsfysize=17cm\epsffile{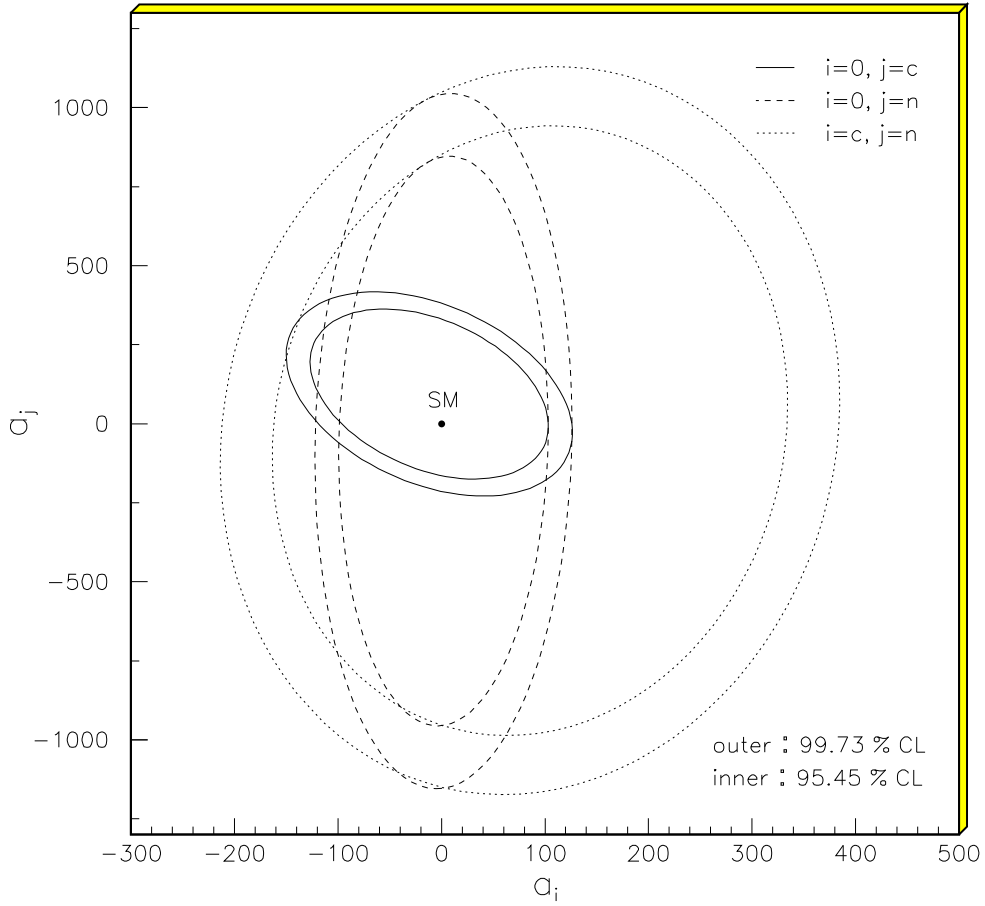}}
\vspace{-2cm}
\caption{\label{ell200}{Contour plots for $+2,+3 \, \sigma$ deviations
from the SM $e^+e^-\to W^+W^-\gamma$ total cross section at $\sqrt{s} =
200$~GeV
with $ \int {\cal L} = 150$~pb$^{-1}$, when two of the three
anomalous couplings  are non-zero.}}
\end{figure}

\begin{figure}[H]
\centerline{\epsfysize=17cm\epsffile{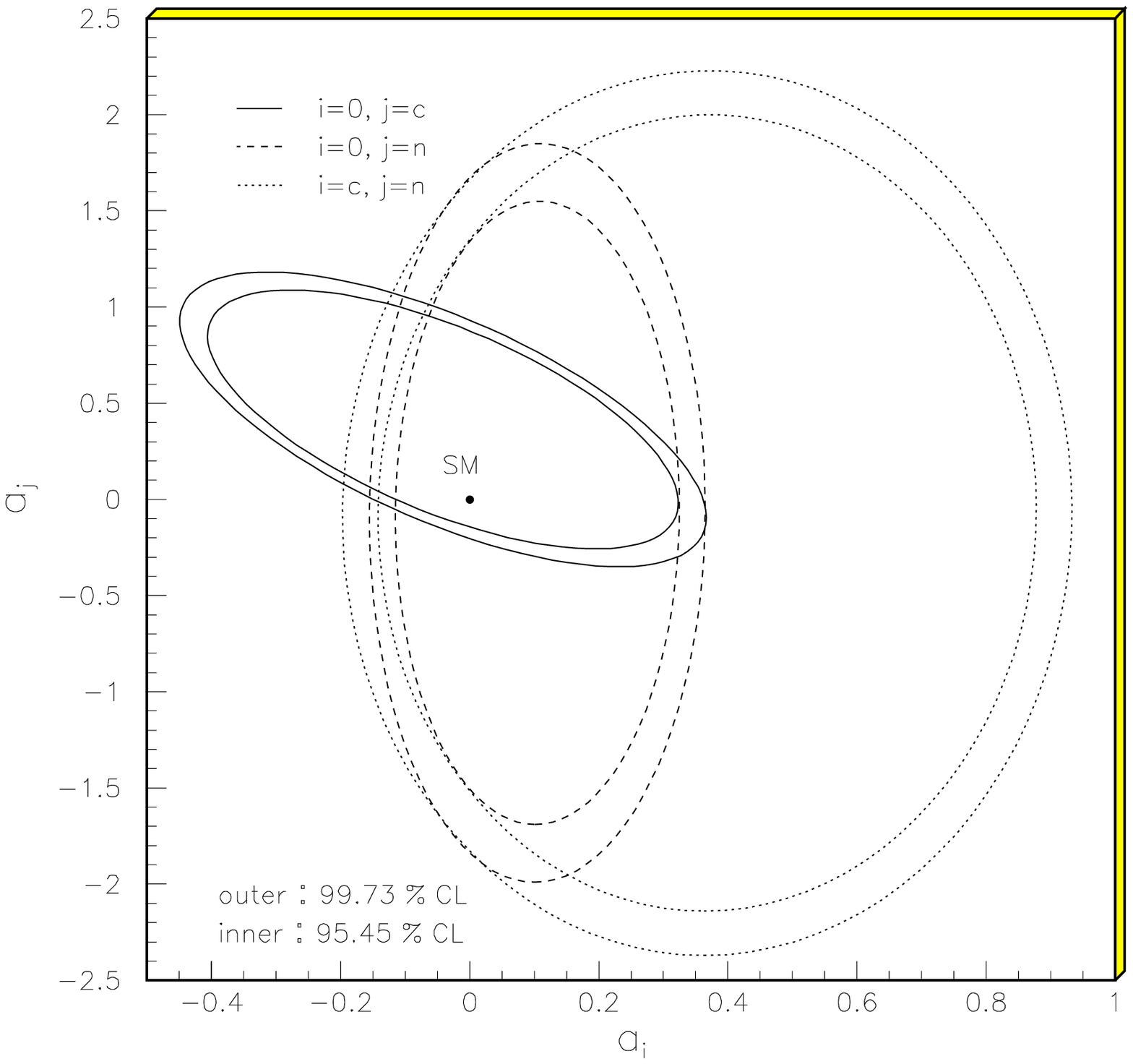}}
\vspace{-2cm}
\caption{\label{ell500}{As for Figure~\ref{ell200}, but for  $\sqrt{s} =
500$~GeV
with $\int {\cal L} = 300$~fb$^{-1}$.}}
\end{figure}

We consider two experimental scenarios: unpolarised $e^+e^- $ collisions
at $200$~GeV with $\int {\cal L} = 150$~pb$^{-1}$, and
at $500$~GeV with ${\cal L} = 300$~fb$^{-1}/\,$year\footnote{In the
following we use the expected integrated luminosity for a run of one year
\cite{TESLA}.}.
Starting with the $W^+W^-\gamma$ process, 
Figure~\ref{ell200} shows the contours in the $(a_i,a_j)$ plane that
correspond
to  $+ 2,+3\,\sigma$ deviations from the SM cross section at $\sqrt{s} =
200$~GeV.
Note that  there are three ellipses, one
for each combination of the three anomalous couplings.
Evidently the sensitivity to $a_0$ and $a_n$ is comparable, corresponding
to $a_i < {\cal O}(100)$ for this luminosity. The corresponding
limit on $a_c$ is some three to four times larger.
Figure~\ref{ell500} shows the same  contours but now at $500$~GeV. The
dramatic improvement
in sensitivity (now $a_i < {\cal O}(1)$) comes partly from the higher
collision
energy (which allows for more energetic photons) but mainly from the much
higher
luminosity. A correlation between the effects of $a_0$ and $a_c$ (solid
ellipses)
is  noticeable at this energy.
\vspace{-1.5cm}
\vspace{-1.5cm}

\begin{figure}[H]
\centerline{\epsfysize=17cm\epsffile{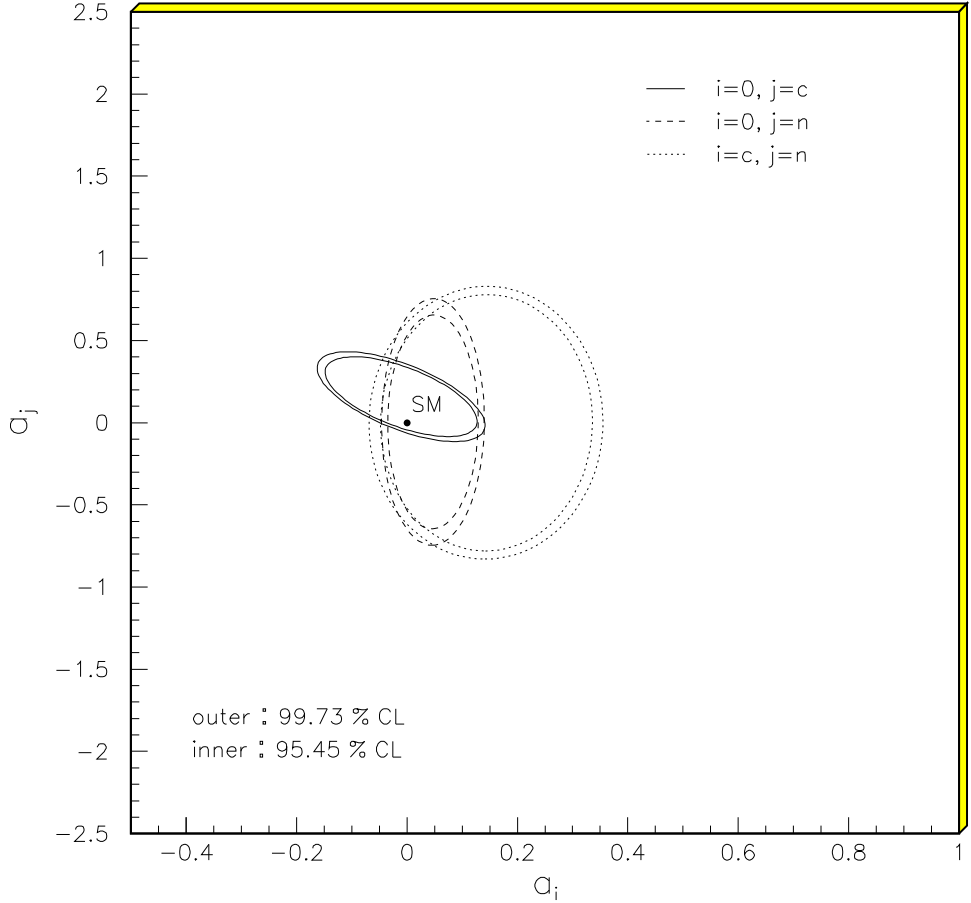}}
\vspace{-2cm}
\caption{\label{rell500}{As for Figure~\ref{ell500}, but with 100\% beam
polarisation.}}
\end{figure}

We have already anticipated a significant improvement in sensitivity for
this process
when the beams are polarised. Specifically, with right-handed electrons
(and left-handed positrons) we suppress a large number of SM `background'
diagrams
where the $W^\pm$ are attached to the fermion line. The effect of 100$\%$
beam polarisation
of this type is shown in Figure~\ref{rell500}.  Assuming the {\it same} 
luminosity
we obtain a factor of approximately 3 improvement in the sensitivity
to an individual anomalous coupling.
\vspace{-1.5cm}

\begin{figure}[H]
\centerline{\epsfysize=17cm\epsffile{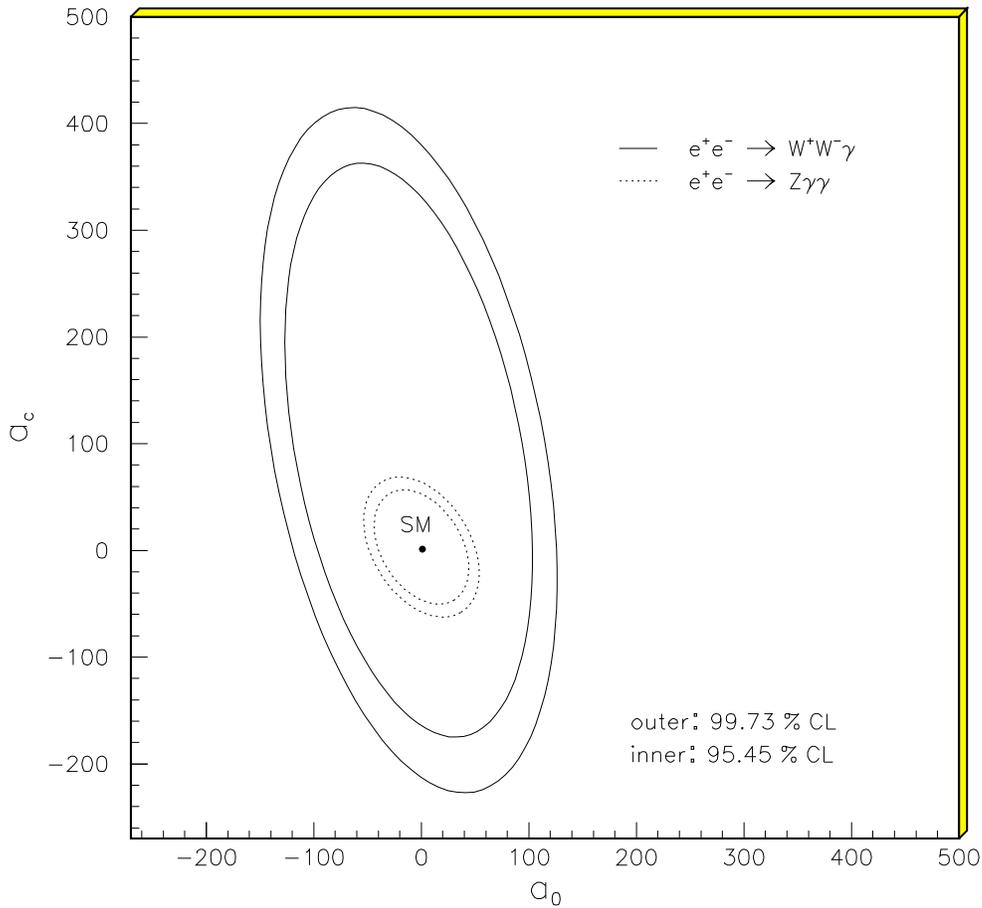}}
\vspace{-2cm}
\caption{\label{com200}{Contour plots for $+2,+3 \, \sigma$ deviations
from the SM 
$e^+e^-\to Z^0\gamma\gamma$
total cross section at $\sqrt{s} = 200$~GeV
with $ \int {\cal L} = 150$~pb$^{-1}$. For comparison, the corresponding
contours for
the $e^+e^-\to W^+W^-\gamma$ process from Figure~\ref{ell200} are also
shown.}}
\end{figure}

Turning to the sensitivity of the $Z^0Z^0\gamma$ and $Z^0\gamma\gamma$
processes,
Figure~\ref{com200} shows the sensitivity of the latter to $a_0$ and $a_c$
at
$\sqrt{s} = 200$~GeV with $\int {\cal L} = 150$~pb$^{-1}$ and unpolarised
beams.\footnote{With
our choice of photon cuts ($E_\gamma > 20$~GeV)
 $\sigma(Z^0Z^0\gamma)$  is essentially zero at this collision energy.}
For comparison, we also show the corresponding $W^+W^-\gamma$ contours
from 
Figure~\ref{ell200}.  The $Z^0\gamma\gamma$ process gives
 a significant improvement in sensitivity, particularly for $a_c$. Since
the SM
 cross sections at this energy are comparable (see Figure~\ref{total}),
the improvement comes
 mainly from the enhanced sensitivity of the matrix element to the
anomalous
 couplings in the $Z^0\gamma\gamma$ case.  
\vspace{-1.5cm}

\begin{figure}[H]
\centerline{\epsfysize=17cm\epsffile{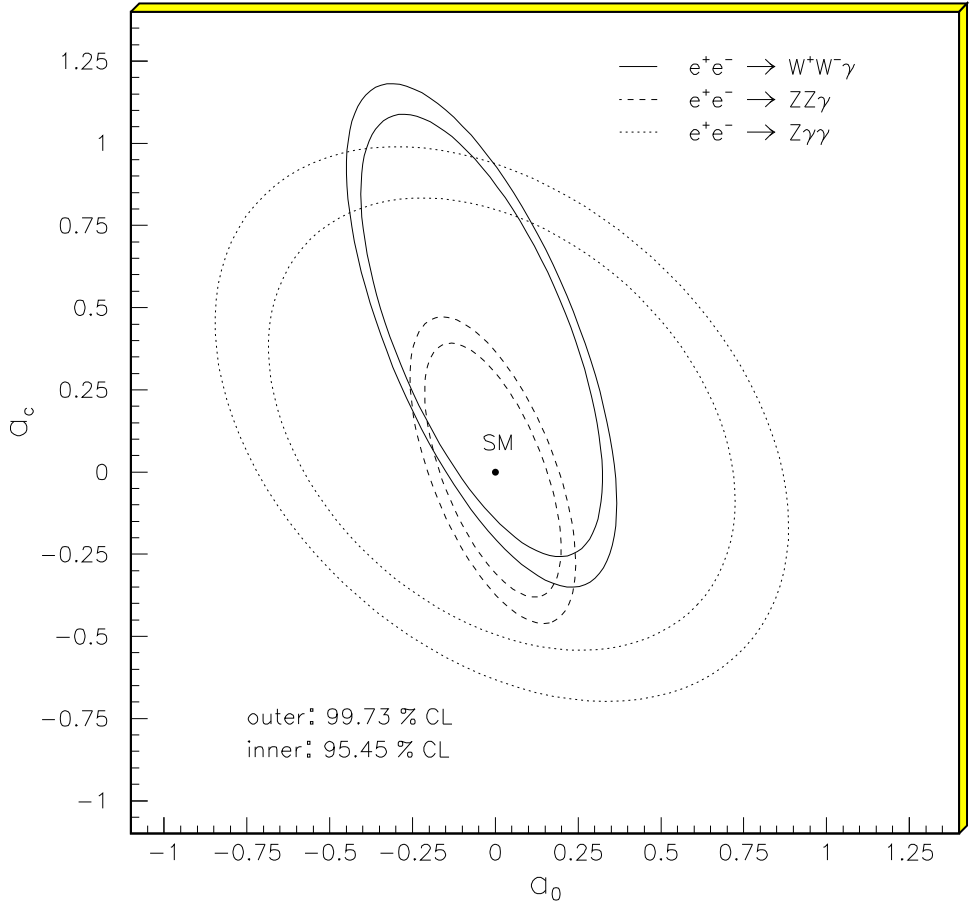}}
\vspace{-2cm}
\caption{\label{com500}{As for Figure~\ref{com200}, but for  $\sqrt{s} =
500$~GeV
with $\int {\cal L} = 300$~fb$^{-1}$ and including also the 
$e^+e^- \to Z^0 Z^0 \gamma$ process.}}
\end{figure}

Finally, Figure~\ref{com500} compares the sensitivity of all three
processes
to $a_0$ and $a_c$ at $\sqrt{s} = 500$~GeV with 
$\int {\cal L} = 300$~fb$^{-1}$ and unpolarised beams. 
The best sensitivity is now provided by the
$Z^0Z^0\gamma$ process
(particularly for $a_c$), despite the fact that it has the smallest cross
section of all the three processes.
Note that polarising the beams has little effect on the sensitivity of the 
$Z^0Z^0\gamma$ and $Z^0\gamma\gamma$ processes to the anomalous couplings,
since the left-handed and right-handed couplings of the $Z^0$ to the
electron
are similar.
\vspace{-1.5cm}

\section{Discussion and Conclusions} 
We have investigated the sensitivity of the processes $e^+e^- \to
W^+W^-\gamma ,\, Z^0Z^0\gamma$ and $Z^0\gamma\gamma$ to genuine anomalous
quartic couplings $(a_0,a_c,a_n)$ at the canonical centre-of-mass energies
$\sqrt{s}=200$~GeV (LEP2) and $500$~GeV (LC). Key features in
determining the sensitivity  for a given process and collision energy, apart
from the fundamental process dynamics,  are the
available photon energy $E_{\gamma}$, the ratio of anomalous diagrams to
SM `background' diagrams, and  the polarisation state of the weak bosons
\cite{belanger}. \\

At $\sqrt{s}=200$~GeV the process $e^+e^- \to Z^0\gamma\gamma$ leads 
to the tightest bounds on the contour of $(a_0,a_c)$, while the process
$e^+e^- \to W^+W^-\gamma$ is needed to set bounds also on $a_n$.
Note that the contours of $(a_0,a_n)$ and $(a_c,a_n)$ can then be improved
using the knowledge of the tighter bounds on  the contour of $(a_0,a_c)$
from $Z^0\gamma\gamma$ production. At this energy $Z^0\gamma\gamma$ benefits
kinematically  from
producing only one massive boson, which leaves more energy for the photons
as well as having fewer `background' diagrams. On the other hand
$W^+W^-\gamma$ production at this energy suffers from the lack
of phase space available for energetic photon emission, although this 
is partially compensated by the production  
of longitudinal bosons, which  gives rise to higher sensitivity
to the anomalous couplings. \\

At $\sqrt{s}=500$~GeV, the effects mentioned above conspire in a somewhat
different way.  All
three processes are now well above their threshold, and hence the availability
of phase space for energetic photons is less of an issue.
 The importance of the longitudinal
polarisation of the massive bosons increases and even though the same number
 of diagrams
contributes to $ Z^0Z^0\gamma$ production as to $Z^0\gamma\gamma$
production, far tighter bounds on the anomalous couplings
can be expected from the former process.
The production of longitudinally polarised bosons is comparable in the
 $ W^+W^-\gamma$ and $ Z^0Z^0\gamma$ processes, but the higher signal 
 to background ratio for the latter leads to a better sensitivity 
 to $a_0$ and $a_c$.\footnote{Here again $W^+W^-\gamma$ is still
needed for investigating $a_n$.} \\

The ability to polarise the beams leads to a significant improvement 
in the sensitivity of the $W^+W^-\gamma$ process, since about a third
of the contributing diagrams are removed. With polarised beams the tightest
bounds now come from this process. The sensitivity of the $e^+e^- \to
Z^0Z^0\gamma$ process is  hardly affected by beam polarisation. 
Furthermore,  for the typical (large) luminosities expected at future
linear colliders \cite{TESLA} the
magnitude of the total cross section itself plays a less important role.\\

The $500$~GeV comparison emphasises the importance of the
longitudinal polarisation states of the massive bosons ($ Z^0Z^0\gamma$ and
$Z^0\gamma\gamma$ are more or less comparable otherwise). This suggests
that the  $e^+e^- \to W^+W^-Z^0$ process should be  more sensitive to anomalous
couplings than $e^+e^-\to  W^+W^-\gamma$, since all three final-state
 particles can be longitudinal polarised. 
 With the expected linear collider luminosity, the somewhat
  smaller cross section should not be an  issue, and  the ratio of background
to signal diagrams is the same as for $W^+W^-\gamma$ production.
Unfortunately this process is only sensitive to $a_n$.\footnote{The $a_0$
and $a_c$ couplings stem from the $VV\gamma\gamma$ vertex.} Furthermore, since
there is no photon in the final state  4-dimensional operators can
also contribute to anomalous couplings (i.e. an anomalous $W^+W^-Z^0Z^0$
vertex) and the analysis becomes significantly more complicated. \\


Finally it is important to emphasise that in our study we have only considered `genuine' quartic couplings from new six-dimensional operators. We have assumed that all other anomalous couplings are zero, including the trilinear
ones.  Since the number of possible couplings and correlations 
is so large, it is in practice very difficult to do a combined analysis of {\it all} couplings simultaneously. In fact, it is not too difficult to think
of new physics scenarios in which effects are only manifest in the quartic
interactions. One example would be a very heavy excited W resonance
produced and decaying as in $W^+ \gamma \to  W^* \to  W^+ \gamma$.\\ 

In principle, any non-zero
trilinear coupling could affect the limits obtained on the quartic
couplings. For example, in equation (4) we showed explicitly
how a non-zero trilinear coupling ($\lambda$) can generate an anomalous
$WW\gamma\gamma$ quartic interaction to compete with the `genuine' ones 
that we have considered. The (dimensionless) strength of the former is $e g  \lambda$, while
for the latter it is $e^2 a_i \langle E_{ext.} \rangle \langle E_{int.} \rangle  / \Lambda^2$, where $ E_{ext.}$ and $E_{int.}$ are the typical
energy scales of the photons entering the vertex. (Here we are considering, as a specific example, the $e^+e^- \to W^+W^-\gamma$ process.)
Since $\Lambda = M_W$, $ \langle E_{ext.} \rangle \sim 25$~GeV and $E_{int.} \sim [5 \sqrt{s} +4(\sqrt{s}- \langle E_{ext.} \rangle)]/9 \sim 190$~GeV , both for $\sqrt{s}=200$~GeV, we see immediately that
the relative contributions of the two types of couplings are in the
approximate  ratio $3 \lambda : a_i$.  Now, at LEP2 upper limits on trilinear couplings
like $\lambda$ are already ${\cal O}(0.1)$ \cite{tgvexpt}. In contrast, we have shown that the limits achievable
on the $a_i$ are ${\cal O}(100)$. Hence we already know that 
the anomalous trilinear couplings have a minimal impact on our analysis.             
The same argument holds at higher collider energies. The limits on the
trilinear couplings will always be so much smaller than those on the quartic
couplings, that they can safely be ignored in studies of the latter.\\

\vspace{0.5cm}
\noindent{\bf Acknowledgements}\\ 
This work was supported in part by the EU Fourth Framework Programme
`Training and Mobility of 
Researchers', Network `Quantum Chromodynamics and the Deep Structure of
Elementary Particles', 
contract FMRX-CT98-0194 (DG 12 - MIHT). AW gratefully acknowledges
financial support in the form of a
`DAAD Doktorandenstipendium im Rahmen des gemeinsamen Hochschulprogramms
III f\"ur Bund und L\"ander'.\\


\end{document}